# Carrier-induced ferromagnetism in n-type ZnMnAlO and ZnCoAlO thin films at room temperature

XH Xu*, AJ Behan, M Ziese**, HJ Blythe, JR Neal, A Mokhtari, MR Ibrahim, AM Fox and G A Gehring

Department of Physics and Astronomy, University of Sheffield, Sheffield S3 7RH

* Permanent address: School of Chemistry and Materials Science, Shanxi Normal University, Linfen, Shanxi 041004, P.R. China

** Permanent address: Division of Superconductivity and Magnetism, University of Leipzig, D-04103 Leipzig, Germany

**Abstract**

**The realization of semiconductors that are ferromagnetic above room temperature will potentially lead to a new generation of spintronic devices with revolutionary electrical and optical properties. Transition temperatures in doped ZnO are high but, particularly for Mn doping, the reported moments have been small. We show that by careful control of both oxygen deficiency and aluminium doping the ferromagnetic moments measured at room temperature in n-type ZnMnO and ZnCoO are close to the ideal values of $5\mu_B$ and $3\mu_B$ respectively. Furthermore a clear correlation between the magnetisation per transition metal ion and the ratio of the number of carriers to the number of transition metal donors was established as is expected for carrier induced ferromagnetism for both the Mn and Co doped films. The dependence of the magnetisation on carrier density is similar to that predicted for the transition temperature for a dilute magnetic semiconductor in which the exchange between the transition metal ions is through the free carriers.**

## I  Introduction

The magnetism observed in dilute magnetic semiconductors (DMS) depends on the existence of delocalised carriers or impurity states that can hybridise with the localised spins on the transition metal sites. The most widely-studied system is GaMnAs, in which the magnetic Mn ions act as a p-type dopant, leading to hole-mediated superexchange [1,2]. While III-V materials serves as a paradigm for understanding the mechanisms of ferromagnetism in semiconductors, their practical usefulness is limited by their low Curie temperatures, which presently lie well below 300 K. In this context, ZnO-based materials have recently been attracting much attention, following the experimental observation of room-temperature ferromagnetism in films doped with a range of transition metal ions [3-9]. Since the band gap of ZnO lies in the ultraviolet at 3.4 eV, these results open up exciting possibilities for spin-optoelectronic devices throughout the entire visible spectral region.

The interest in ZnO was originally prompted by theoretical predictions concerning hole-mediated magnetism, as in the III-V systems [10]. However, the experimental work has been almost entirely concerned with *n-type* materials, which raises important and interesting scientific issues concerning the carrier-mediated magnetism. In this article we present a detailed study of the effects of controlled n-type doping of ZnMnO and ZnCoO thin films by incorporation of aluminium. This has enabled us to investigate the relationship between the magnetism and the electron density in a systematic way. This is in contrast to previous work in which the doping level was typically controlled by introducing defects through an oxygen deficiency. We find a clear relationship between the carrier density and the magnetism, which has enabled us to obtain strongly enhanced magnetic moments, especially for the ZnMnO films.

The results for the Mn-doped system are particularly significant. The $Mn^{2+}$ ion has the largest magnetic moment, $5\mu_B$, of any of the transition elements, but so far the measured magnetic moments per Mn ion have been disappointingly small. Sharma et al [3] found a moment of 0.16 $\mu_B$ and Coey et al [4] found a moment that was so much smaller than that for other elements such as Co or Ti that they classed it as nonmagnetic. The moment per atom was increased dramatically by Norberg et al [11] who found a moment of 1.35 $\mu_B$ in thin films made by depositing nanocrystals on a surface; these authors used a very low concentration of 0.2%, giving a small total magnetization (they reported that the magnetisation fell rapidly for doping concentrations above 0.5% and was essentially zero at 2.5%[7]). In this article we show that the moment on Mn at 2% doping is increased dramatically in an *n*-type sample to a maximum of ~4.5 $\mu_B$ at room temperature by co-doping with Al to provide electrons as carriers. This moment is comparable with the best moments that have been obtained in GaMnAs [2] but with the important difference that it occurs at room temperature.

Another aspect of ZnMnO that makes it particularly interesting is that it alone, among the ZnMO series [M= transition metal], has been shown to be magnetic when p-type [7,8,11]. Combined with these observations from other groups, our results lead to the exciting possibility of producing a strongly magnetic pn junction, which is beyond the present capability of magnetic III-V semiconductors.

**II Sample Preparation and Structural Data**

Stoichiometric targets of composition $Zn_{1-x-y}Al_xMn_yO$ with (y = 0.02) were prepared by mixing appropriate amounts of 99.999% pure ZnO, $MnO_2$ and $Al_2O_3$ with subsequent grinding for about 15 min. The powders of Mn and Al co-doped ZnO were fired at $400^0C$ and the pressed pellet was calcined at $400^0C$ for 8 h. The resulting pellets were used as targets in a pulsed laser deposition (PLD) chamber, where they were ablated using a Lambda Physik XeCl laser ($\lambda$= 308 nm) at a repetition rate of 10 Hz. During deposition, the partial oxygen pressure was held constant at a value between base pressure ($3\times10^{-5}$ Torr) and 1 Torr. The films, with thickness varying from 60 to 1000 nm, were grown either on R-cut or C-cut sapphire substrates held at temperature $450^0C$. The sample thickness was measured using a Dektak profilometer; film structure was characterized by X-ray diffraction; magnetization measurements were carried out in a Quantum Design SQUID magnetometer; Hall effect and magneto-resistance measurements were made in a continuous flow cryostat equipped with an iron bore magnet. The experimental results are summarized in Table I including the thickness data although no systematic dependence of the magnetization was observed with film thickness.

X-ray diffraction (XRD) patterns of the $Zn_{0.94}Co_{0.05}Al_{0.01}O$ film and target sample are shown in figure 1a and in figure 1b for a $Zn_{0.97}Mn_{0.02}Al_{0.01}O$ film. The powder XRD patterns indicate that the target is single phase with a hexagonal wurtzite structure. Traces of Al and Co metal, their respective oxides, or any binary zinc cobalt phases were not observed in the target sample. The lattice constants measured from XRD of the target sample are a = 0.3227nm and c = 0.5143nm which are smaller than those of pure bulk ZnO (a = 0.3250nm and c = 0.5207nm). The decrease in the lattice parameters is consistent with the substitution of $Zn^{2+}$ ions (ionic radius in tetrahedral configuration 74 pm) by the smaller $Co^{2+}$ (72 pm) and $Al^{3+}$ (53 pm) ions. Only the (110) intensity peak of the ZnO film and the reflections from the substrate appear in the XRD pattern. This indicates that the R-cut sapphire substrate induces highly oriented growth of the co-doped ZnO film. The lattice parameter a =0.3237nm is also smaller than that of pure ZnO. The corresponding XRD pattern of the $Zn_{0.97}Mn_{0.02}Al_{0.01}O$ film, made under the same deposition conditions, again shows only the (110) reflex, indicating highly oriented growth on R-cut sapphire. Whereas many studies report an increase of the lattice parameter with increasing Mn content, the film here has a lattice constant very close to that of pure ZnO. This is attributed to the incorporation of $Al^{3+}$ ions in the lattice which have a considerably smaller ionic radius than both $Zn^{2+}$ and $Mn^{2+}$ ions (80 pm).

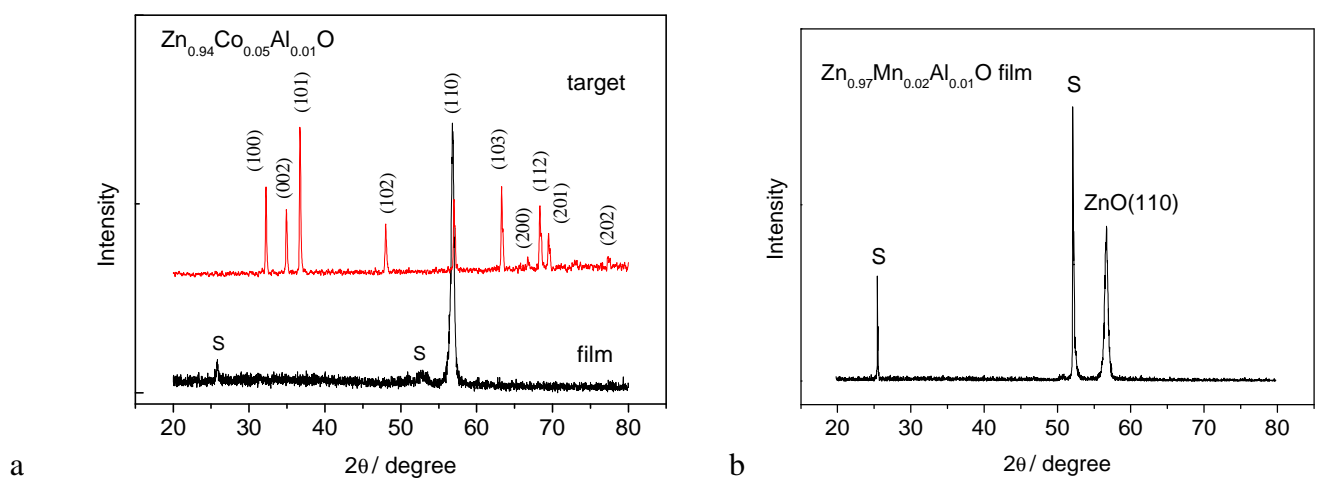

Figure 1. XRD patterns of the (a) $Zn_{0.94}Co_{0.05}Al_{0.01}O$ film and target and (b) $Zn_{0.97}Mn_{0.02}Al_{0.01}O$ film

# III Measurements of the magnetisation

The results for the magnetic moments for the films grown under different conditions are shown in Table I.

**TABLE I**: A summary of the data on the samples shown in the plots. The labels are marked on the figures. All samples were grown on R-cut sapphire substrates, except for those marked * which were grown on C-cut sapphire. The first column gives the concentrations of the dopants, the second the oxygen pressure in the PLD chamber, the thickness is given (we did not observe a systematic dependence of the magnetization on thickness), the magnetization is the saturation moment, in Bohr magnetons per transition metal ion, measured in the SQUID at room temperature. The final column gives the carrier density measured at room temperature.

| Label | TM and Al concentration | $O_2$ (mTorr) | Thickness (nm) | $M_s$ ($\mu_B$ / TM) 300 K | $n_c$ (cm$^{-3}$) 300 K |
|---|---|---|---|---|---|
| 1 | 5% Co | 0.06 | 280 | 0.19 | 8.5E+19 |
| 2 | 5% Co + 1% Al | 200 | 260 | 0.00 | 3.4E+18 |
| 3 | 5% Co + 1% Al | 0.06 | 160 | 1.69 | 7.1E+20 |
| 4 | 5% Co | 50 | 280 | 0.00 | 9.3E+19 |
| 5 | 5% Co + 1% Al | 25 | 280 | 0.63 | 3.0E+20 |
| 6 | 5% Co | 0.06 | 120 | 0.88 | 4.3E+19 |
| 7 | 5% Co + 1% Al | 0.3 | 150 | 1.31 | 8.4E+20 |
| 8 | 5% Co + 1% Al | 15 | 170 | 1.17 | 5.7E+20 |
| 9 | 5% Co | 0.3 | 220 | 0.79 | 5.6E+20 |
| 10 | 5% Co + 1% Al | 50 | 360 | 0.70 | 2.1E+20 |
| 11 | 5% Co + 2% Al | 0.06 | 290 | 0.16 | 1.5E+21 |
| 12* | 5% Co + 1% Al | 0.06 | 120 | 0.68 | 9.7E+20 |
| 13* | 5% Co + 1% Al | 10 | 100 | 0.13 | 1.8E+21 |
| a | 2% Mn + 1% Al | 50 | 310 | 1.90 | 6.7E+20 |
| b | 2% Mn + 2% Al | 0.06 | 1150 | 0.13 | 1.4E+21 |
| c | 2% Mn + 2% Al | 0.6 | 1270 | 0.13 | 1.1E+21 |
| d | 2% Mn + 2% Al | 60 | 330 | 0.54 | 1.8E+21 |
| e | 2% Mn + 4% Al | 0.06 | 360 | 0.48 | 1.2E+21 |
| f | 2% Mn + 2% Al | 0.06 | 570 | 2.32 | 9.5E+20 |
| g | 2% Mn + 1% Al | 0.06 | 400 | 4.36 | 6.3E+20 |
| h | 2% Mn + 1% Al | 50 | 320 | 3.68 | 2.8E+20 |
| i* | 2% Mn | 10 | 80 | <0.10 | 1.4E+18 |
| j* | 2% Mn | 5 | 120 | <0.10 | 1.8E+18 |
| k* | 2% Mn | 15 | 120 | <0.10 | 6.4E+17 |
| m* | 2% Mn | 10 | 580 | <0.10 | 3.0E+19 |
| n* | 2% Mn + 1% Al | 0.06 | 530 | 4.25 | 1.4E+21 |
| o* | 2% Mn + 1% Al | 0.06 | 230 | 0.12 | 1.4E+21 |
| p | 2% Mn + 1% Al | 300 | 300 | 2.4 | 1.1E+20 |
| x* | 1% Al | 10 | 100 | 0 | 2.5E+21 |
| y* | 1% Al | 0.06 | 250 | 0 | 1.1E+21 |
| z* | - | 10 | 330 | 0 | 8.8E+18 |

Figure 2 shows an example of the room temperature SQUID magnetisation data of the cobalt doped samples, with and without additional aluminium, as a function of the oxygen pressure in the deposition chamber. The magnetisation values given in Table I are the saturation values obtained from such plots. The data has been corrected for the diamagnetic contribution of the substrate.

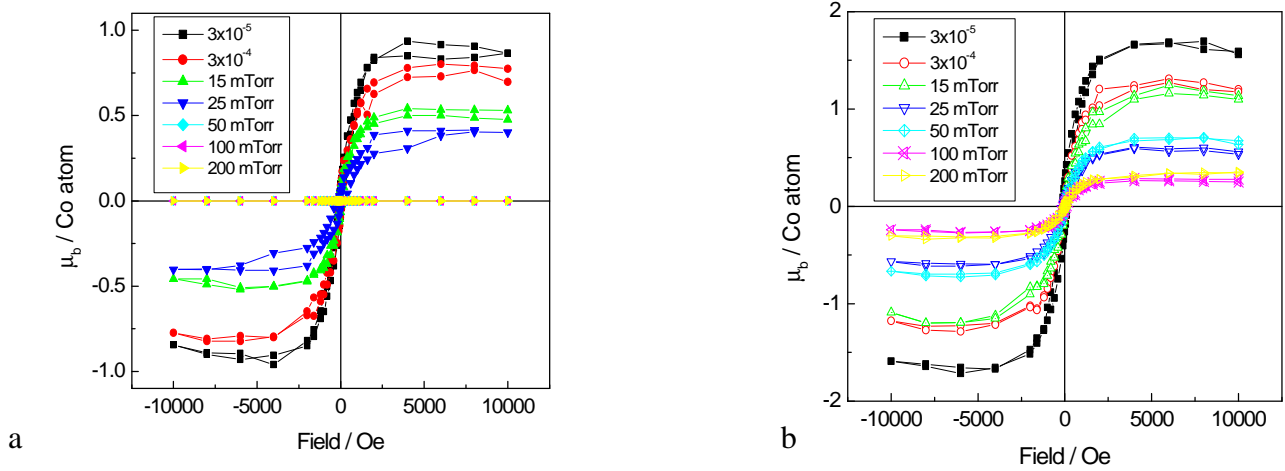

Figure 2. Hysteresis loops of the (a) $Zn_{0.95}Co_{0.05}O$ and (b) $Zn_{0.95}Co_{0.04}Al_{0.01}O$ films. All the measurements were made at room temperature.

Figure 3 shows the saturation magnetisation measured at room temperature plotted as a function of the oxygen pressure in the deposition chamber. For ease of cross-referencing each data point is identified with an entry in Table I.

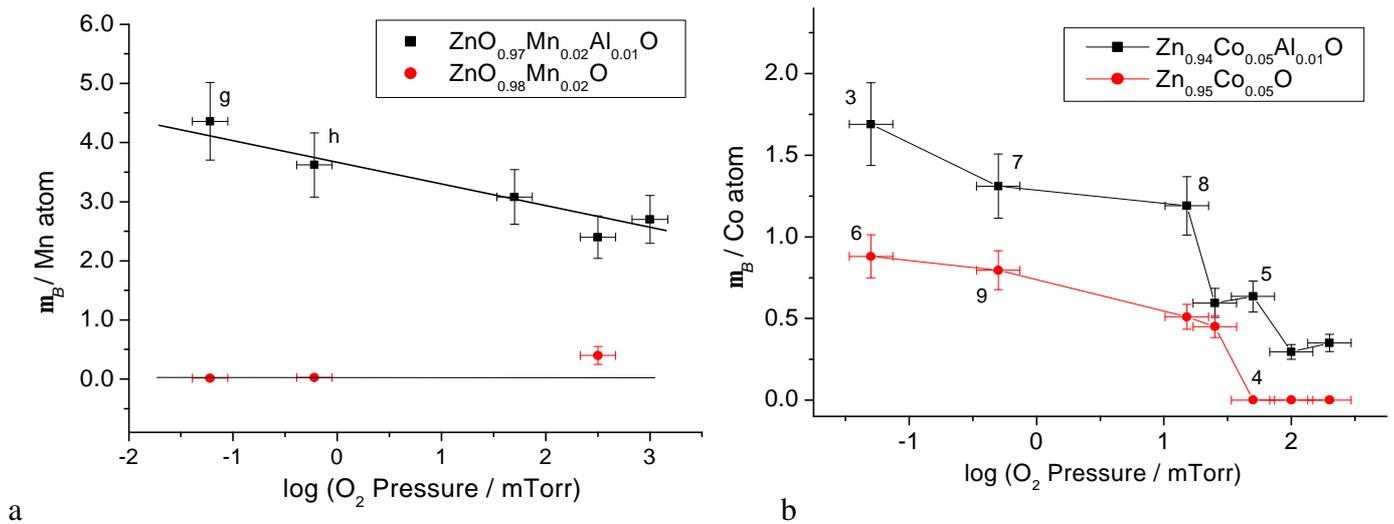

Figure 3. Room temperature magnetization of the (a) $Zn_{0.98}Mn_{0.02}O$, $Zn_{0.97}Mn_{0.02}Al_{0.01}O$ and (b) $Zn_{0.95}Co_{0.05}O$, $Zn_{0.94}Co_{0.05}Al_{0.01}O$ films as a function of the oxygen pressure. Other information on these films is collected in Table I.

These plots make it clear that we obtain the best films when they are grown at low pressure close to the base pressure which is 0.03mTorr in our chamber. In figure 4 we present the magnetisation data as a function of aluminium doping for films all grown at base pressure. These show that, for films grown at this pressure, there is a dramatic increase in the magnetisation for 1% doping of aluminium. The results for Mn doping show that the magnetic moment is qualitatively different for films that contain aluminium.

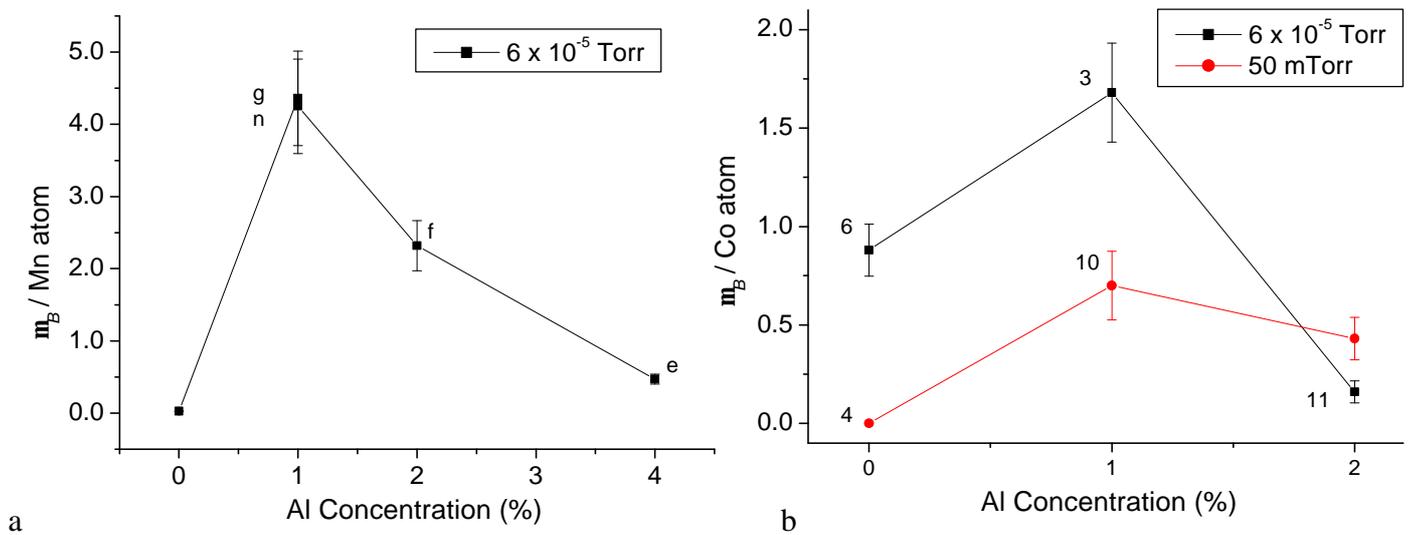

Figure 4. Room temperature magnetization of (a) ZnMnO and (b) ZnCoO films as a function of the Al concentration. Other information on these films is collected in Table I.

**IV Discussion of the Results**

A characteristic of DMS is that an increased carrier density enhances both the magnetisation and the Curie temperature. Free electrons can be introduced into transition-metal-doped ZnO in two ways. The samples can be grown under an oxygen deficiency to produce shallow donor states, usually through Zn interstitials. Alternatively, donors can be deliberately introduced via co-doping with a group III element. In this work, we explore the combination of both methods, with aluminium as the group III dopant. We can obtain insight into the results presented in figures 3 and 4 by considering the effects of the different growth conditions on the free electron density. Films of pure ZnO grow $n$-type due to Zn interstitials or oxygen vacancies. In A PLD system the oxygen non-stoichiometry may be controlled by adjusting the oxygen pressure in the chamber [12] and Zn interstitials may be added later by exposing a film to Zn vapour [13]. For a film produced by PLD, with substrate temperature $T_s = 750\,°C$ the carrier density and the mobility both peak at an oxygen pressure of $10^{-3}$ Torr [12]. Films grown at lower pressure, e.g. $10^{-5}$ Torr, have fewer carriers at room temperature although they are expected to have a higher density of donor levels associated with the oxygen deficiency. We note that the results for ZnCoO in figure 1b, which agree with those of Coey *et al.* [3], demonstrate the link between the magnetic moment and the oxygen pressure, and hence, indirectly, between the magnetic moment and the carrier density.

The effect of Al-doping depends on the other defects present. The carrier concentration of the ZnCoO and ZnMnO films measured at room temperature as a function of the oxygen pressure in the PLD growth chamber are presented in figure 4. In the absence of Al donors there is a strong difference between the carrier densities of ZnO doped with Mn or Co. The plots are of the ratio of the number densities of free carriers to the magnetic ions because this is the ratio that is of interest in DMS [2].

The carrier density for Mn-doped samples without Al is $\sim 10^{17}$ cm$^{-3}$ is similar to that for pure ZnO [12] whereas the carrier density for ZnCoO is much higher $\sim 10^{20}$ cm$^{-3}$ [9]. The high carrier density of ZnCoO is due to the hybridization of the Co d states with the donor levels, and results in its more robust magnetic properties [3,7]. However, when the films are doped with Al the carrier concentrations of the Mn and Co doped films are comparable, as can be seen from Table I.

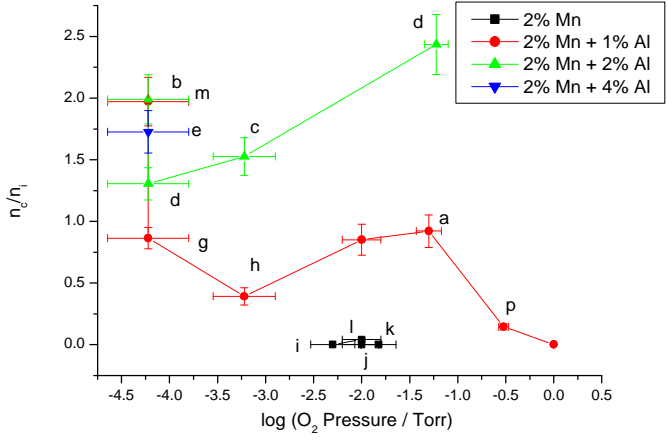
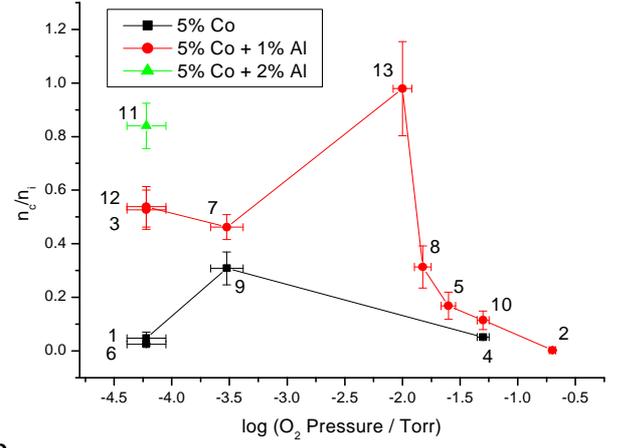

Figure 5. The ratio of the carrier density, $n_c$, to the density of magnetic ions, $n_i$, in (a) 2% Mn film and (b) 5% Co doped film with and without Al doping as a function of oxygen pressure. The numbers refer to the samples in Table I.

The variation of carrier concentration with oxygen pressure for ZnCoO qualitatively tracks that for pure ZnO peaking at an oxygen pressure of ~1mTorr. However when 1 % Al is added the carrier concentration no longer falls at lower oxygen pressure. The Mn doped films have very low carrier densities in the absence of any Al. Indeed the addition of 5% Mn has been shown to reduce the leakage current by suppressing the number of free electrons in highly insulating ZnO films [14]; this is consistent with the results that show that Mn can act as an acceptor [7,8].

In figure 6 we plot the magnetisation as a function of the ratio of the carrier density to the density of magnetic ions for the Mn and Co doped samples. Figure 6a shows that the magnetic moments for the Mn doped samples is much higher when Al is co-doped but that the moment does fall as the ratio $n_c/n_i$ rises above unity. We note that the 1% Al samples have high magnetization when they appear to contribute one carrier each ($n_c/n_i$ ~0.5). It is clear that the free carrier density is important as expected for DMS, but that the electrons are not the only mediators of exchange because of the important contribution from the donor states [15]. What is new here is that when the donor level is high enough the hybridization of the Mn levels with the conduction band occurs and actually gives even stronger magnetism than occurred on the p-type materials and at a concentration of 2% compared with 0.2% as required for the p-doped sample.

An interesting feature of the result is that magnetization depends on the density of the free carriers and is relatively insensitive to whether they arose from Al doping or by oxygen nonstoichiometry. The transition metal ions will contribute to the ferromagnetism provided that the interaction with the donor or conduction electrons is stronger than that due to the antiferromagnetic interactions. Superexchange between transition metal ions that are near neighbours through oxygen ions is much stronger than occurs through arsenic which means that the competing antiferromagnetic interactions are much more important in doped ZnO than in doped GaAs [16]. Hence the magnetisation measured at room temperature may be used as a measure of the strength of the ferromagnetic exchange. If the ferromagnetic superexchange is able to dominate over all the antiferromagnetic coupling except that due to nearest neighbours then one would expect that the observed moment, $m$, at concentration $x$ would result from all the transition metal ions that did *not* have a transition metal ion as a nearest neighbour be related to the ideal moment, $m_0$, by $m = m_0(1-x)^{z_{eff}}$ where $z_{eff} \sim 12$ is the number of nearest neighbours assuming a random distribution. This theory predicts values of the moment for 2% Mn and 5% Co samples of $3.9\mu_B$ and $1.6\mu_B$ respectively. Such values are close to the maximum observed values presented in figures 1 and 2. Fitzgerald et al [17] developed a more sophisticated cluster model which included the moments from larger clusters to fit the dependence of the magnetisation of a Co doped sample and deduced that there was some clustering in their samples. There may also be contributions to the moments from the oxygen donor levels and the conduction electrons.

The dependence of the transition temperature on the density of mobile carriers is regarded as good evidence for a genuine DMS [1,2,18]. A similar relationship should exist between the room temperature magnetic moment per transition metal ion and the transition temperature. This follows because both effects depend on the strength of the induced exchange between the transition metal ions.
For free carrier mediated magnetism the transition temperature is maximized when the ratio of the carrier density to the number of free ions is ~0.5 and falls off quite rapidly when the ratio exceeds unity [18]. This prediction is borne out by the magnetisation data presented in figure 5, the curve which is a guide to the eye follows the theoretical curve for the transition temperature for carrier induced ferromagnetism [18].

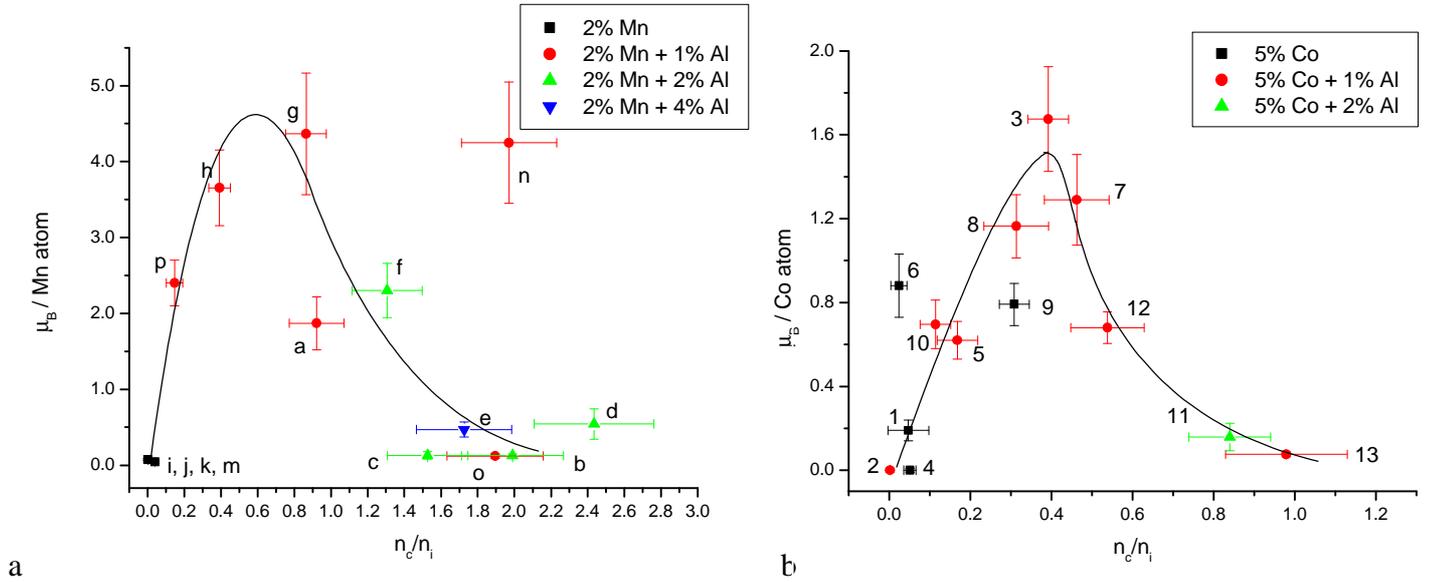

Figure 6. Plots of the room temperature magnetisation per TM ion for (a) Mn (2%) and (b) Co (5%) doped in ZnO as a function of the ratio of the carrier density also measured at room temperature to the density of magnetic ions. The numbers refer to the samples in Table I. The curve is a guide to the eye.

The effects of spin orbit coupling are far smaller in *n*-type doped ZnO compared with p-type Mn doped GaAs because the conduction bands are formed from *s* electrons. This has several consequences. The anomalous Hall effect (AHE) that is large in GaMnAs is unobservable in ZnO [19,20] because the AHE varies at least linearly with the spin-orbit coupling [2]. The anisotropic magneto-resistance, which varies as the square of the spin orbit coupling, is unobservable in ZnO.

However since the magneto-resistance (MR) depends on the exchange between the localized and delocalised electrons, it may be large even in the absence of spin orbit interactions. This is another indication of the interactions between the free carriers and the localized spins. Both negative [19,21] and positive [22] values for the MR have been reported.

We find a very strong dependence of the MR on carrier density and no measurable anisotropy. In figure 7a we show the MR taken at 4K for ZnO, ZnMnO and ZnCoO samples all of which have low carrier densities. We see that the MR is negative for ZnO and positive for magnetically doped films.

At 4K a negative MR occurs in heavily doped semiconductors and in figure 7b we show the MR for Al doped films all of which have similar, high, carrier densities. The scale is different for this plot as the effects are all much smaller. All the MR is negative but that for pure ZnO has the largest negative MR thus indicating again that the inclusion of magnetic ions makes a small positive contribution to the overall negative signal. Paramagnetic ions are expected to give a negative MR as the applied field will align the spins and so reduce the scattering. A possible cause of the positive MR in the ferromagnetic materials is an increase in the band splitting in a field that will reduce the overall density of states.

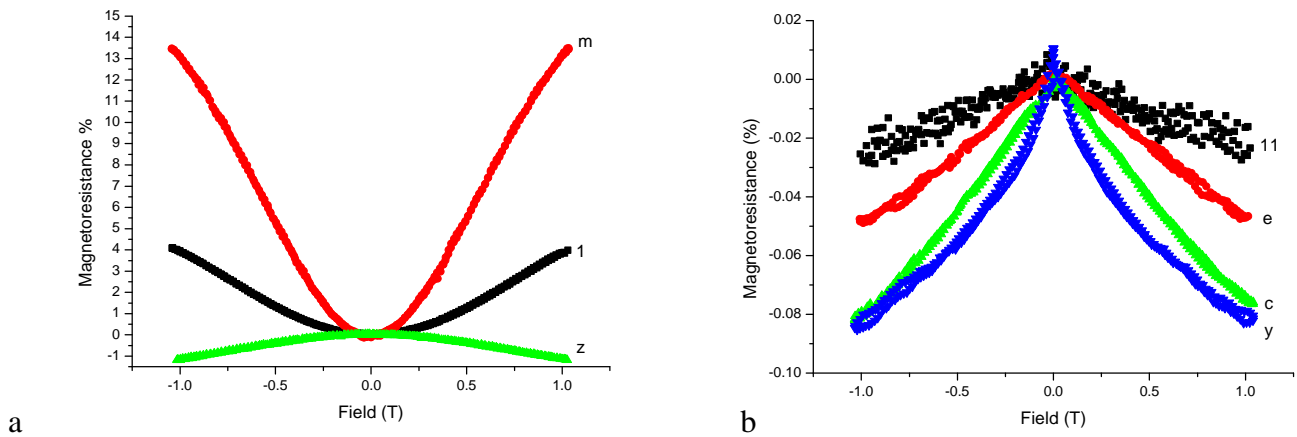

Figure 7. The MR loops at 4K for (a) samples with low carrier density and (b) comparison of the MR of samples with high (and comparable) carrier density and different magnetic properties. Note the difference in scale the maximum MR value shown in figure 7b is –0.09% compared with +13.5% in figure 7a. The sample numbers relate to the data in Table I.

**Conclusion**

The main scientific issue that is addressed in our results is the effect of the free electron density on the magnetism. The most obvious point to make is that the inclusion of free electrons by Al-doping can be highly beneficial for the magnetism in ZnO. This clearly indicates the importance of electron-mediated exchange, which contrasts with the hole-mediated mechanism that prevails in GaMnAs and also exchange through the hybridisation of the transition metal ions with the localised donor states.

We have shown that large magnetic moments may be obtained in n-type ZnO by careful control of the donors. This is the first report of a DMS with a room temperature magnetic moment that is close to the ideal value and also has a substantial ($\sim 10^{21}$ cm$^{-3}$) carrier density. We find that the contribution to the exchange from the free carriers arising from oxygen deficiency is surprisingly similar to that arising from Al doping. A very significant result is that we obtained a magnetic moment for Mn doped ZnO that is close to the ideal value. All the magnetization measurements were made at room temperature. The dependence of the magnetization on the carrier density is shown to vary in a way that is predicted by theory for $T_c$. The interaction of the carriers with the local spins is shown by the different MR obtained for ferromagnetic and nonmagnetic samples with similar carrier density. Finally it is clear that doped ZnO is a controllable *n*-type room temperature DMS with very low spin orbit coupling. This makes it interestingly different from GaMnAs which is *p*-type DMS with very strong spin orbit coupling.

## Acknowledgements

The authors are grateful for support from the Engineering and Physical Sciences Research Council (EPSRC) and Royal Society of UK and the National Science Foundation of China. One of us (GAG) is grateful for useful discussions with T Jungwirth and B Gallagher.